\documentclass[12pt,a4paper]{article}

\usepackage{ifthen} 
\newboolean{pdflatex}
\setboolean{pdflatex}{true} 

\newboolean{articletitles}
\setboolean{articletitles}{true} 

\newboolean{uprightparticles}
\setboolean{uprightparticles}{false} 

\usepackage{float}
\usepackage{multicol}
\usepackage{comment}
\usepackage{url}
\usepackage{subcaption} 

\def\paperauthors{ 
M.~Alexander$^{1}$,
W.~Barter$^{2}$,
G.~Bogdanova$^{3}$,
S.~Borghi$^4$,
T.J.V.~Bowcock$^{5}$,
E.~Buchanan$^{6}$,
C.~Burr$^{6}$,
K.~Carvalho~Akiba$^{7}$,
S.~Chen$^{8}$,
P.~Collins$^{6}$,
E.~Dall'Occo$^{7}$,
S.~De~Capua$^{4}$,
C.T.~Dean$^{9}$,
F.~Dettori$^{10}$,
K.~Dreimanis$^{11}$,
G.~Dujany$^{12}$,
R.~Dumps$^{6}$,
L.~Eklund$^{13}$,
T.~Evans$^{4}$,
M.~Ferro-Luzzi$^{6}$,
M.~Gersabeck$^{4}$,
T.~Gershon$^{14}$,
L.~Grillo$^{1,4}$, 
T.~Hadavizadeh$^{15}$,
K.~Hennessy$^{5}$,
W.~Hulsbergen$^{7}$,
D.~Hutchcroft$^{5}$,
E.~Jans$^{7}$,
M.~John$^{16}$,
P.~Kopciewicz$^{17}$,
P.~Koppenburg$^{7}$,
T.~Latham$^{14}$,
A.~Leflat$^{3}$,
M.W.~Majewski$^{17}$,
B.~Mitreska$^{4}$,
A.~Oblakowska-Mucha$^{17}$,
C.~Parkes$^{4}$,
A.~Poluektov$^{18}$,
W.~Qian$^{19}$,
S.~Redford$^{20}$,
K.~Rinnert$^{5}$,
E.~Rodrigues$^{5}$,
G.~Sarpis$^{21}$,
M.~Schiller$^{1}$,
T.~Szumlak$^{17}$,
M.~Tobin$^{8}$,
M.~van~Beuzekom$^{7}$,
V.~Volkov$^{3}$,
M.R.J.~Williams$^{22}$,
} 
  
\def\paperasciititle{Measurement of thermal properties of the LHCb VELO detector using track-based software alignment}  
\def\papertitle{Measurement of thermal properties of the LHCb VELO detector using track-based software alignment} 
\def\paperkeywords{{High Energy Physics}, {LHCb}} 
\def\papercopyright{\the\year\ CERN for the benefit of the LHCb collaboration} 
\def\paperlicence{CC-BY-4.0 licence}
\def\paperlicenceurl{https://creativecommons.org/licenses/by/4.0/}

\usepackage[top=1in, bottom=1.25in, left=1in, right=1in]{geometry}

\usepackage{microtype}
\usepackage{lineno}  
\usepackage{xspace} 
\usepackage{caption} 

\usepackage{graphicx}  
\usepackage{color}
\usepackage{colortbl}
\graphicspath{{./figs/}} 
\DeclareGraphicsExtensions{.pdf,.PDF,png,.PNG}

\usepackage{amsmath} 
\usepackage{amssymb}
\usepackage{amsfonts}
\usepackage{upgreek} 

\newcommand*\patchAmsMathEnvironmentForLineno[1]{%
\expandafter\let\csname old#1\expandafter\endcsname\csname #1\endcsname
\expandafter\let\csname oldend#1\expandafter\endcsname\csname
end#1\endcsname
 \renewenvironment{#1}%
   {\linenomath\csname old#1\endcsname}%
   {\csname oldend#1\endcsname\endlinenomath}%
}
\newcommand*\patchBothAmsMathEnvironmentsForLineno[1]{%
  \patchAmsMathEnvironmentForLineno{#1}%
  \patchAmsMathEnvironmentForLineno{#1*}%
}
\AtBeginDocument{%
\patchBothAmsMathEnvironmentsForLineno{equation}%
\patchBothAmsMathEnvironmentsForLineno{align}%
\patchBothAmsMathEnvironmentsForLineno{flalign}%
\patchBothAmsMathEnvironmentsForLineno{alignat}%
\patchBothAmsMathEnvironmentsForLineno{gather}%
\patchBothAmsMathEnvironmentsForLineno{multline}%
\patchBothAmsMathEnvironmentsForLineno{eqnarray}%
}

\usepackage{hyperxmp}

\usepackage[pdftex,
            pdfauthor={\paperauthors},
            pdftitle={\paperasciititle},
            pdfkeywords={\paperkeywords},
            pdfcopyright={Copyright (C) \papercopyright},
            pdflicenseurl={\paperlicenceurl}]{hyperref}

\usepackage[colorinlistoftodos,textsize=scriptsize]{todonotes}

\usepackage[all]{hypcap} 

\usepackage{xspace} 
\usepackage{upgreek}

\def\lhcb   {\mbox{LHCb}\xspace}

\def\lhc    {\mbox{LHC}\xspace}

\def\velo   {VELO\xspace}

\def\MagUp {\mbox{\em Mag\kern -0.05em Up}\xspace}

\ifthenelse{\boolean{uprightparticles}}%
{

 \def\PDelta      {\ensuremath{\Delta}\xspace}                 
 \def\PXi         {\ensuremath{\Xi}\xspace}                 
 \def\PLambda     {\ensuremath{\Lambda}\xspace}                 
 \def\PSigma      {\ensuremath{\Sigma}\xspace}                 
 \def\POmega      {\ensuremath{\Omega}\xspace}                 
 \def\PUpsilon    {\ensuremath{\Upsilon}\xspace}

 \def\PB      {\ensuremath{\mathrm{B}}\xspace}                 
                  
 \def\PD      {\ensuremath{\mathrm{D}}\xspace}

 \def\PK      {\ensuremath{\mathrm{K}}\xspace}

 \def\Pi      {\ensuremath{\mathrm{i}}\xspace}

 \def\thebaroffset{0.0em}
}
{

 \mathchardef\PDelta="7101
 \mathchardef\PXi="7104
 \mathchardef\PLambda="7103
 \mathchardef\PSigma="7106
 \mathchardef\POmega="710A
 \mathchardef\PUpsilon="7107
                  
 \def\PB      {\ensuremath{B}\xspace}                 
                  
 \def\PD      {\ensuremath{D}\xspace}

 \def\PK      {\ensuremath{K}\xspace}

 \def\Pi      {\ensuremath{i}\xspace}

 \def\thebaroffset{0.18em}
}
\newcommand{\offsetoverline}[2][\thebaroffset]{\kern #1\overline{\kern -#1 #2}}%

\makeatletter
\ifcase \@ptsize \relax
  \newcommand{\miniscule}{\@setfontsize\miniscule{4}{5}}
\or
  \newcommand{\miniscule}{\@setfontsize\miniscule{5}{6}}
\or
  \newcommand{\miniscule}{\@setfontsize\miniscule{5}{6}}
\fi
\makeatother

\DeclareRobustCommand{\optbar}[1]{\shortstack{{\miniscule (\rule[.5ex]{1.25em}{.18mm})}
  \\ [-.7ex] $#1$}}

\def\KorKbar {\kern \thebaroffset\optbar{\kern -\thebaroffset \PK}{}\xspace}

\def\DorDbar {\kern \thebaroffset\optbar{\kern -\thebaroffset \PD}\xspace}

\def\BorBbar {\kern \thebaroffset\optbar{\kern -\thebaroffset \PB}\xspace}

\def\Y#1S{\ensuremath{\PUpsilon{(#1S)}}\xspace}

\def\LorLbar     {\kern \thebaroffset\optbar{\kern -\thebaroffset \PLambda}\xspace}

\def\AT#1     {\ensuremath{A_{\mathrm{T}}^{#1}}\xspace}           

\def\C#1      {\ensuremath{\mathcal{C}_{#1}}\xspace}                       
\def\Cp#1     {\ensuremath{\mathcal{C}_{#1}^{'}}\xspace}                    
\def\Ceff#1   {\ensuremath{\mathcal{C}_{#1}^{\mathrm{(eff)}}}\xspace}        
\def\Cpeff#1  {\ensuremath{\mathcal{C}_{#1}^{'\mathrm{(eff)}}}\xspace}       
\def\Ope#1    {\ensuremath{\mathcal{O}_{#1}}\xspace}                       
\def\Opep#1   {\ensuremath{\mathcal{O}_{#1}^{'}}\xspace}                    

\newcommand{\nospaceunit}[1]{\ensuremath{\text{#1}}}       
\newcommand{\aunit}[1]{\ensuremath{\text{\,#1}}}       

\newcommand{\tev}{\aunit{Te\kern -0.1em V}\xspace}
\newcommand{\gev}{\aunit{Ge\kern -0.1em V}\xspace}
\newcommand{\mev}{\aunit{Me\kern -0.1em V}\xspace}
\newcommand{\kev}{\aunit{ke\kern -0.1em V}\xspace}
\newcommand{\ev}{\aunit{e\kern -0.1em V}\xspace}
\newcommand{\mevc}{\ensuremath{\aunit{Me\kern -0.1em V\!/}c}\xspace}
\newcommand{\gevc}{\ensuremath{\aunit{Ge\kern -0.1em V\!/}c}\xspace}
\newcommand{\mevcc}{\ensuremath{\aunit{Me\kern -0.1em V\!/}c^2}\xspace}
\newcommand{\gevcc}{\ensuremath{\aunit{Ge\kern -0.1em V\!/}c^2}\xspace}

\def\mum  {\ensuremath{\,\upmu\nospaceunit{m}}\xspace}

\def\degc {\ensuremath{^\circ}{\text{C}}\xspace}

\def\muRad {\ensuremath{\,\upmu\nospaceunit{rad}}\xspace}

\def\gsim{{~\raise.15em\hbox{$>$}\kern-.85em
          \lower.35em\hbox{$\sim$}~}\xspace}
\def\lsim{{~\raise.15em\hbox{$<$}\kern-.85em
          \lower.35em\hbox{$\sim$}~}\xspace}

\def\murad  {\ensuremath{\,\upmu\nospaceunit{rad}}\xspace}

\xspace

\def\tell1  {TELL1\xspace}
\def\ukl1   {UKL1\xspace}

\usepackage{cite} 
\usepackage{mciteplus}

\usepackage{longtable} 
\begin{document}

\renewcommand{\thefootnote}{\fnsymbol{footnote}}
\setcounter{footnote}{1}


\begin{titlepage}
\pagenumbering{roman}

\vspace*{-1.5cm}
\centerline{\large EUROPEAN ORGANIZATION FOR NUCLEAR RESEARCH (CERN)}
\vspace*{1.5cm}
\noindent
\begin{tabular*}{\linewidth}{lc@{\extracolsep{\fill}}r@{\extracolsep{0pt}}}
\ifthenelse{\boolean{pdflatex}}
{\vspace*{-1.5cm}\mbox{\!\!\!\includegraphics[width=.14\textwidth]{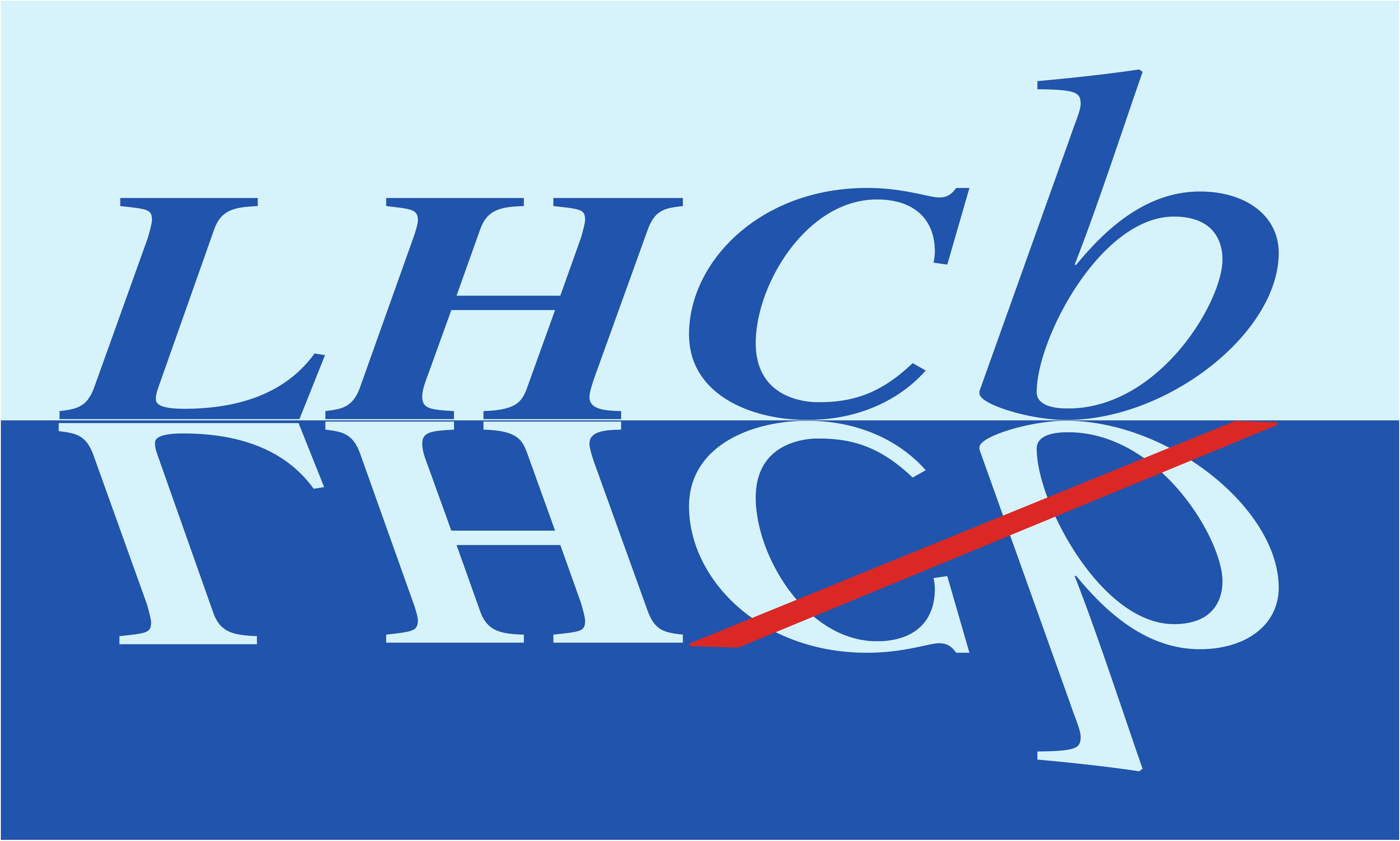}} & &}%
{\vspace*{-1.2cm}\mbox{\!\!\!\includegraphics[width=.12\textwidth]{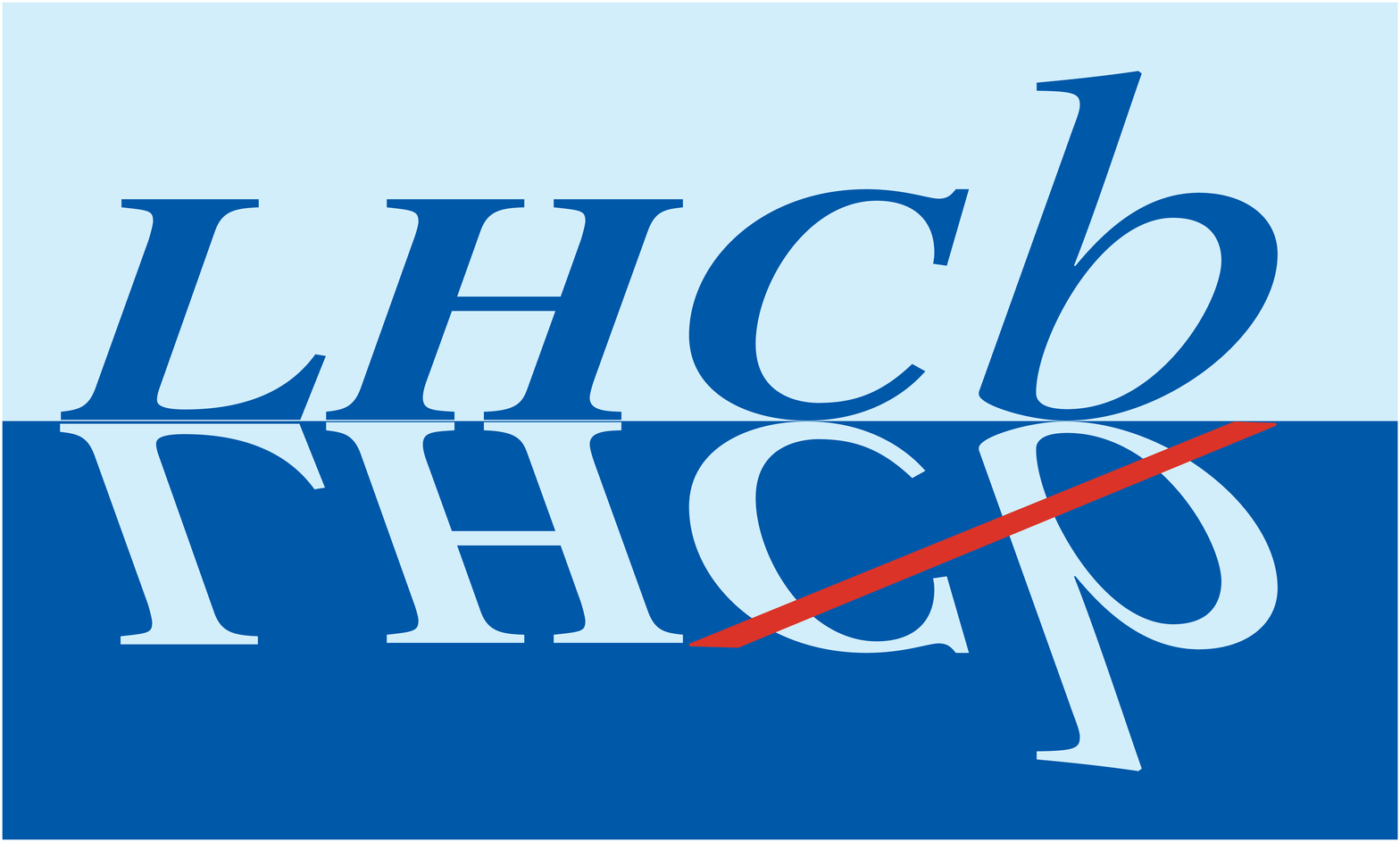}} & &}%
\\
 & & \today \\ 
 & & \\ 
\end{tabular*}

\vspace*{4.0cm}

{\normalfont\bfseries\boldmath\huge
\begin{center}
  \papertitle 
\end{center}
}

\vspace*{2.0cm}

\begin{center}
The LHCb VELO group
\end{center}

\vspace{\fill}

\begin{abstract}
  \noindent
%
%
  The thermal properties of the \lhcb Vertex Locator (\velo) are studied using the real-time detector alignment procedure. The variation of the position and orientation of the detector elements as a function of the operating temperature of the \velo is presented. This study uses a dataset collected by the LHCb experiment during a \velo temperature scan performed at the end of LHC Run 2 (October 2018). Significant shrinkage of the \velo modules is observed at the operating temperature of $-30$\degc compared to the laboratory measurements on a single module taken at a range of temperatures from +45\degc to -25\degc. The thermal shrinkage expected from the extrapolation of laboratory measurements to lower temperatures, and the results of this alignment study are in good agreement.
  
\end{abstract}

\vspace*{2.0cm}

\begin{center}
  Submitted to JINST 
\end{center}

\vspace{\fill}

{\footnotesize 
\centerline{\copyright~\papercopyright. \href{\paperlicenceurl}{\paperlicence}.}}
\vspace*{2mm}

\end{titlepage}


\newpage
\setcounter{page}{2}
\mbox{~}
%
%
%
%

\cleardoublepage


\renewcommand{\thefootnote}{\arabic{footnote}}
\setcounter{footnote}{0}



\pagestyle{plain} 
\setcounter{page}{1}
\pagenumbering{arabic}
 

%


\newpage

\section{Introduction}
\label{sec:Introduction}
 
The \lhcb experiment \cite{lhcb-exp} was designed to investigate the difference between matter and antimatter by studying decays of beauty and charm hadrons. It is located at the LHC (Large Hadron Collider) situated at CERN. Precision measurements of the properties of $b$ and $c$ hadron decays and searches for rare decays help understanding the Standard Model and are among the most promising strategies to search for new particles and couplings. As the production of the $b$-nadrons is concentrated in the forward direction, the \lhcb experiment was constructed as a single arm forward spectrometer which covers the pseudorapidity region of $ 2< \eta < 5$. The closest detector to the proton-proton interaction region is  the Vertex Locator (\velo). The \velo is a silicon detector that reconstructs trajectories of the particles passing through and distinguishes between primary and secondary vertices. The tracking system reconstructs the particles' trajectories (tracks) and determines their momentum. The RICH (Ring Imaging Cherenkov) detectors provide particle identification using Cherenkov radiation. A dipole magnet curves the paths of charged particles and allows the momenta to be determined. Their energy is measured by an electromagnetic and hadronic calorimeter. The last sub-detector of the experiment is the muon system which detects muons and measures their properties. 
The trigger system of the experiment determines which events are recorded and selected to be analysed. From the recorded information, different variables for the decay particles can be calculated and used for further data selection and analysis.

\begin{figure}[H]
  \begin{center}
    \includegraphics[width=0.5\linewidth]{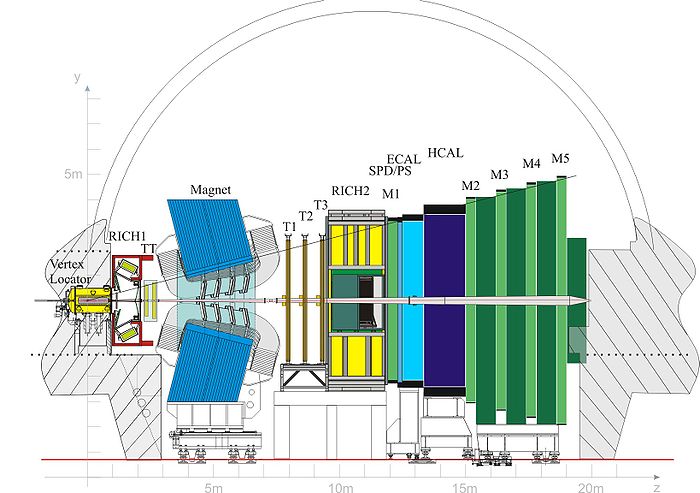}
  \end{center}
  \caption{Side view of the \lhcb experiment. Figure taken from \cite{lhcb-exp}.} 
  \label{fig:Lhcb}
\end{figure}

This paper presents the \velo module position stability as a function of the operational temperature variation. The ideal tool for this study is the alignment procedure applied to a sample of reconstructed tracks, which is used to detect translations and rotations of detector elements. 

\section{Design of the \velo for \lhc Runs 1 - 2 (2010 - 2018)}

\begin{figure}[H]
\centering
\begin{subfigure}{.5\textwidth}
  \centering
  \includegraphics[scale=0.3]{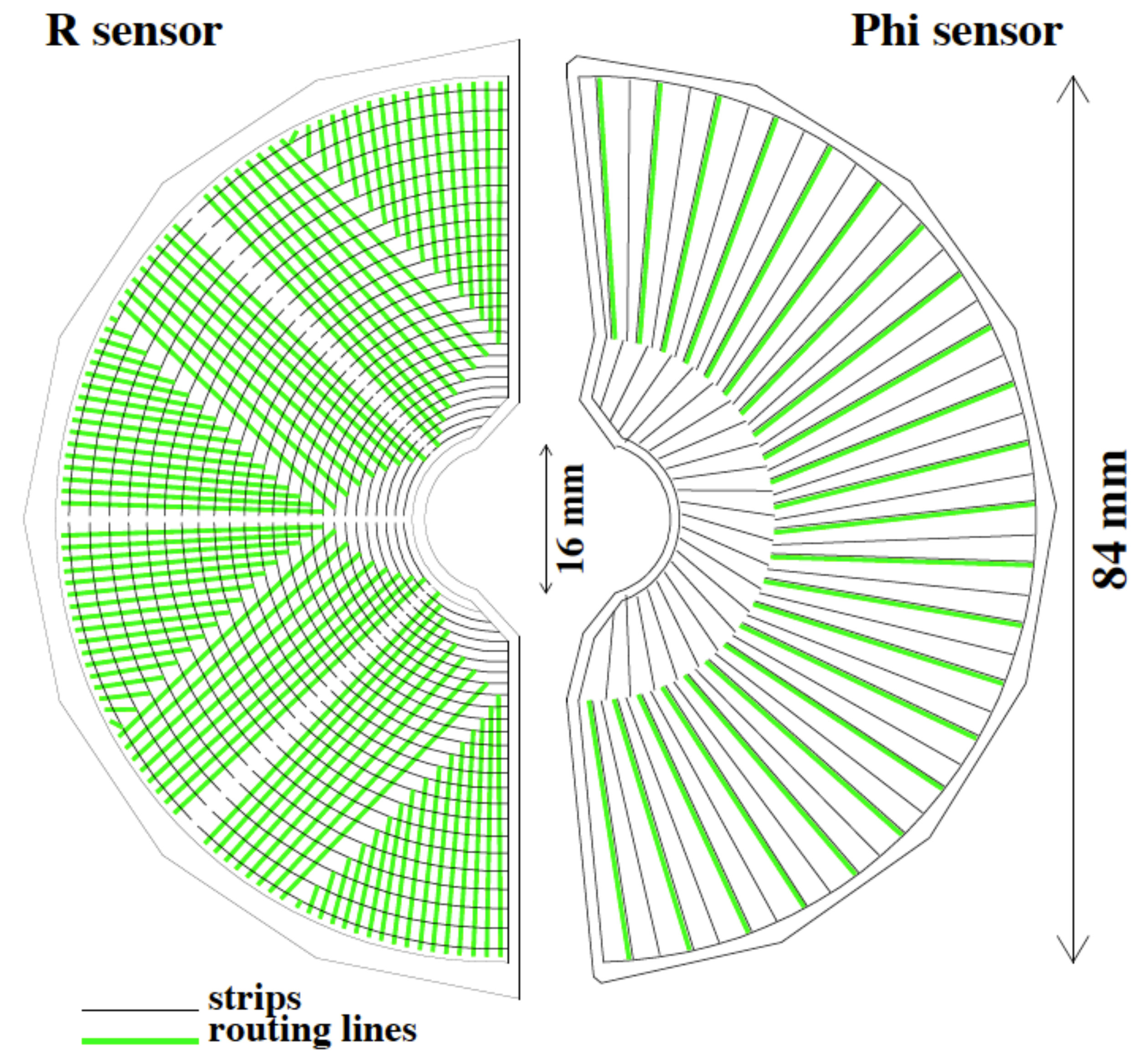} 
  \caption{}
  \label{fig:r-phi-sensor}
\end{subfigure}%
\begin{subfigure}{.5\textwidth}
  \centering
  \includegraphics[scale=0.5]{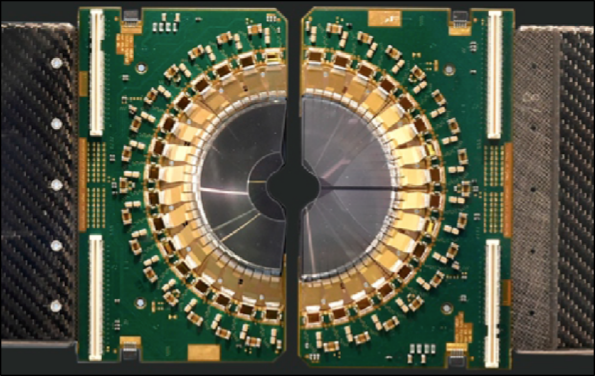} 
  \caption{}
    \label{fig:mod-run2}
\end{subfigure}
\caption{Schematic representation of an $R$ and a $\phi$ sensor (a) and a photograph of the \velo strip detector module (b). Figures taken from \cite{LHCb-DP-2014-001} and \cite{Buchanan:2243156}.} 
\end{figure}

The \velo has the role of locating primary vertices, defined as the proton-proton interaction points, and secondary vertices from decays of long lived particles. The Run 1 - 2 \velo design \cite{LHCb-TDR-005} consists of 42 silicon modules arranged perpendicularly to the beam operating at a set temperature of $-30$\degc (the set point of the cooling system). 
The \velo modules each provide a measurement of $R$ and $\phi$ coordinates using the 84 single-sided radial ($R$) and azimuthal-angle ($\phi$) strip sensors. The $R$ and $\phi$ sensors are mounted on both sides of a highly thermally conductive spine which supports the readout hybrid, and the modules are supported with a carbon fibre paddle stand. The $R$ sensor strips are arranged into four approximately 45$^\circ$ segments and have routing lines perpendicular to the strips. The $\phi$ sensor has two zones with inner and outer strips. The routing lines of the inner strips are orientated parallel to the outer strips (schematically depicted in Fig.~\ref{fig:r-phi-sensor} and photographed in Fig.~\ref{fig:mod-run2}). 

\begin{figure}[H]
\centering
\begin{subfigure}{.5\textwidth}
  \centering
  \includegraphics[height=.5\linewidth]{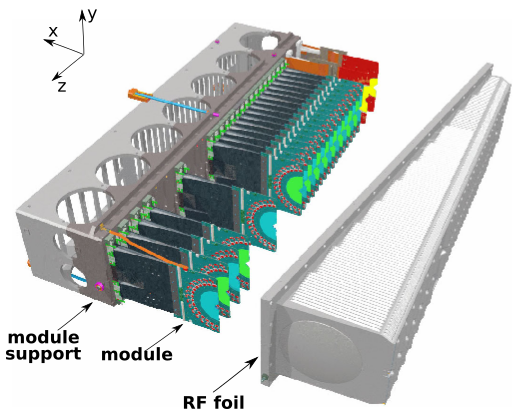}
  \caption{}
  \label{fig:velo-1} 
\end{subfigure}%
\begin{subfigure}{.5\textwidth}
  \centering
  \includegraphics[height=.5\linewidth]{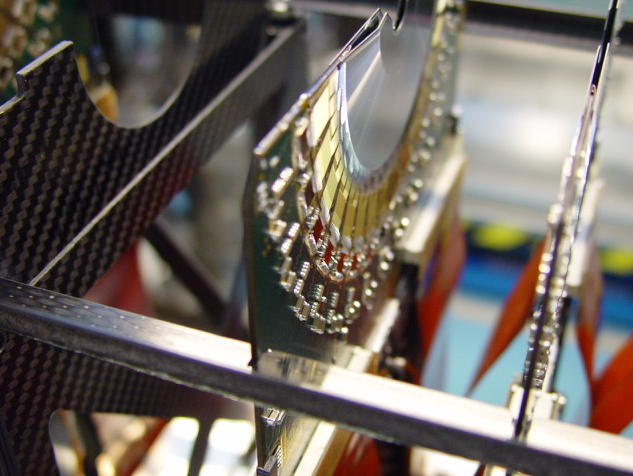} 
  \caption{}
    \label{fig:constraint-system}
\end{subfigure}
\begin{subfigure}{.5\textwidth}
  \centering
  \includegraphics[height=.8\linewidth]{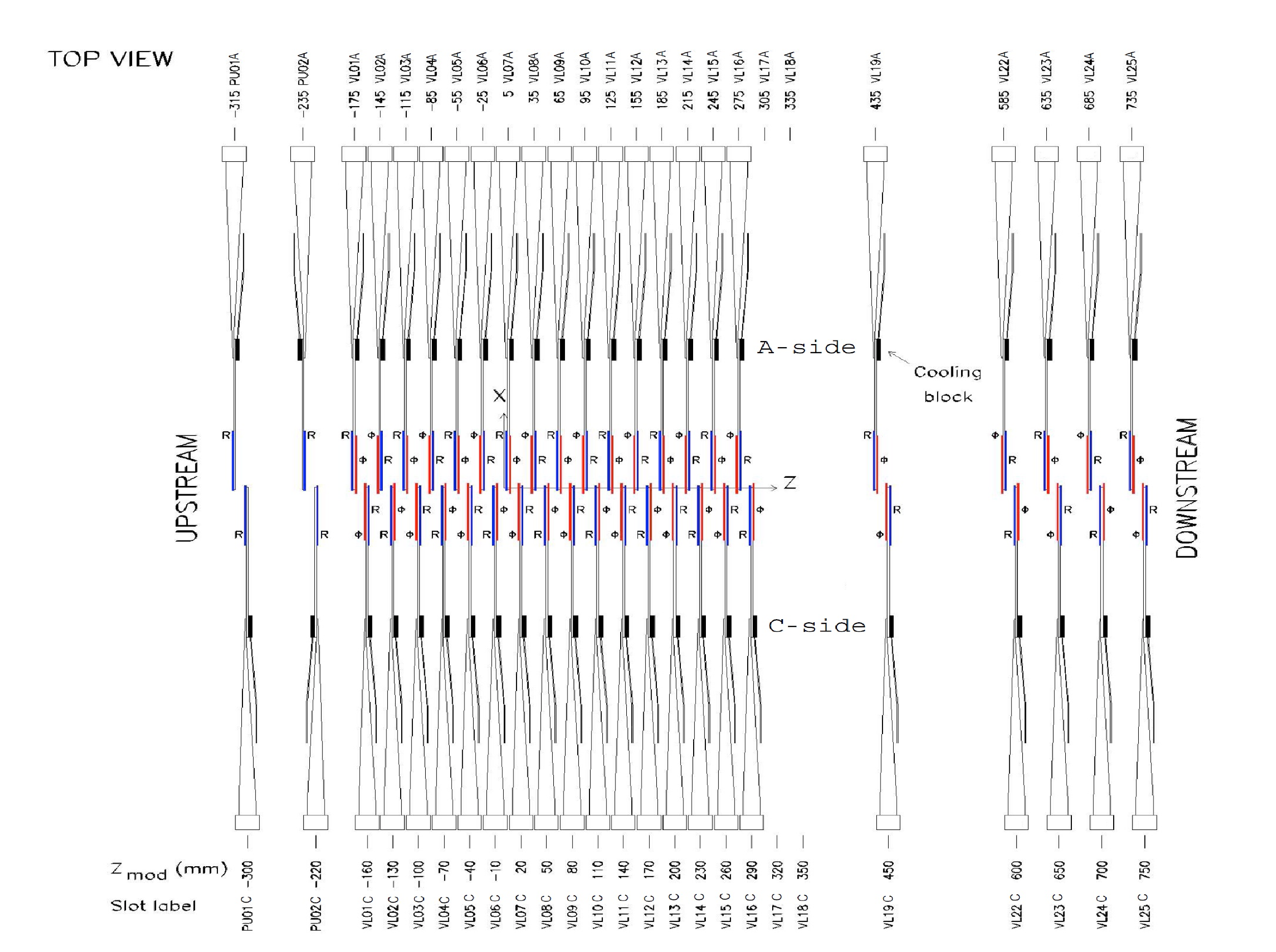} 
  \caption{}
    \label{fig:velo-layout}
\end{subfigure}
\caption{The Run 1-2 \velo design. One of the two removable halves of the detector is shown with the beam axis placed along the z direction (a) \cite{Buchanan:2243156}. A photograph of one of the slots of the \velo constraint system (b). The \velo module layout along the $z$ direction (c).}
\end{figure}

The \velo modules are arranged perpendicularly to the beam along a length of about 1~m (see Fig.~\ref{fig:velo-1}). Their layout along the $z$ direction, same as the beam direction, is shown on Fig.~\ref{fig:velo-layout}. The module number takes the odd numbers on the right half and the even numbers on the left half of the detector (important later for the alignment plots). The temperature of the detector is maintained constant via a cooling system that uses mixed-phase $\rm CO_2$ that is supplied as a liquid into a number of stainless steel pipes embedded within aluminium pads which are clamped to the base of the module. The mechanical and thermal link to the sensors is provided by the carbon fibre support. As result of the temperature variation during the cooling the Copper (Cu) cables effectively pull the module along $z$ \cite{lhcb-exp}, hence this movement had to be limited. This was achieved by installing a $z$-constraint system. The constraint system is composed of precisely designed slots (about 100 $\mum$ wider than the thickness of the modules) in which the modules were inserted, shown on Fig.~\ref{fig:constraint-system}. This ensured the modules remained around their nominal $z$ position. The whole detector operates in secondary vacuum conditions separated from the \lhc vacuum by a thin aluminium foil (Fig. \ref{fig:velo-1}). 
The two halves are retracted at $\pm$ 29 mm from the LHC beam for safety during the beam injection and beam operation. When stable beam is declared, the halves are closed with an accuracy of 10\mum around the beam at an approximate distance of 8 mm.
This \velo detector operated from October 2009 up to the end of \lhc Run 2 in December 2018. It is being replaced by a new detector in Run 3. 

\section{The \velo alignment}

The performance of the \lhcb \velo has ensured track reconstruction efficiency above 98$\%$, a vertex resolution along the beam axis of about 70~\mum, and a decay time resolution of about 45~fs \cite{LHCb-DP-2014-001}. One of the key ingredients to achieve such performance is the alignment. The correct \velo alignment contributes to correctly distinguishing primary and secondary vertices and to improve the quality of the impact parameter. The alignment of the \velo is determined by a Kalman filter \cite{Kalman-alignment, Amoraal:2012qn} track model that is used in a minimum-$\chi^2$ algorithm to determine the position of the detector elements (halves, modules and sensors) from a sample of reconstructed tracks. The position of each detector element is stored in a database that is accessed by the software for reconstruction. The alignment algorithm estimates the misplacement of a set of positions of detector elements with respect to their expected positions after closure of the \velo. The software aligns for three translations ($\rm T_x, T_y, T_z$) and three rotations ($\rm R_x, R_y, R_z$) along/around the $x$, $y$ or $z$ axis of the \velo which are known as alignment constants. 

For the Run~2 data-taking period the \lhcb experiment developed a new data taking strategy \cite{LHCB-DP-2019-001} including the full-detector real-time alignment and calibration procedure \cite{Realtime}. During data taking, when the \lhc beams are declared stable, the halves of the \velo move towards the beam and the \velo is closed. The closing and opening of the \velo repeats for each fill of the \lhc, which can lead to alignment changing in each fill. The alignment has to be run at the beginning of the fill in order to check if the detector elements have the expected position and orientation, which assures high track reconstruction performance and unbiased vertex position reconstruction. If there is a significant difference between the alignment constants calculated by two consecutive alignments, the most recent values are used to update the alignment.  
The whole alignment procedure can run over several iterations until it converges. A convergence is reached if a certain condition for the $\chi^2$ of tracks is fulfilled. The alignment minimizes the total $\chi^2$ of a sample of tracks, with the $\chi^2$ defined as 
\begin{equation}
    \chi^2_{tot} = \sum_{track}^{tracks}\chi^2_{track}.
\end{equation}
\noindent
The tracks are fitted with the standard LHCb Kalman filter track fit, which includes corrections from multiple scattering. In the LHCb algorithm for the VELO alignment the $\chi^2$ receives additional terms from vertex constraints \cite{Amoraal:2012qn} and terms to account for the knowledge of detector positions from survey.
The algorithm allows to apply the survey constraints at different levels of granularity, which in case of the \velo are the surveyed positions of modules in the halves, and the positions of the halves in the global \lhcb frame. The assigned uncertainty of the VELO module survey in the VELO half frame is 20\mum for translations and 200\murad for rotations. 
The alignment uses reconstructed tracks and imposes selection on the track quality. A requirement of the number of hits and cuts on momentum and $\chi^2$ of tracks are applied. Along with the tracks, the reconstructed primary vertices are used as an input to the \velo alignment. A selection that contains the maximum and minimum number of tracks of the vertex and requirement on the $\chi^2$ for the vertex fit is used. 
To assess the convergence of the alignment the total predicted change in the $\chi^2$ is considered, which is defined as
\begin{eqnarray}
    \delta\chi^2 = \alpha^T \rm{Cov}(\alpha^{-1}) \alpha,
\end{eqnarray}

\noindent
where $\alpha$ is the change in the alignment parameters. In the minimization procedure the set of alignment parameters is translated into a set of independent alignment modes by diagonalizing the covariance matrix. The alignment is considered converged when both the total $\frac{\delta \chi^2}{\rm ndof} < 4$, and for each alignment mode $\delta \chi^2_i < 25$. 
The alignment needs to perform an update of the position of the detector elements if there has been a significant variation during data-taking.
A variation is considered significant when the difference between the value of the alignment constant determined in the most recent alignment, and the previous value, exceeds the ``min variation'' values reported in Table~\ref{tab:tab1}. In order to reduce the sensitivity to statistical effects, the alignment parameters used for data processing are only updated if the variation determined by the alignment algorithm is considered significant. Any variation below the ``min variation" threshold is considered as fluctuation within the accuracy of the alignment. The constants are updated every few fills and overall good stability is found~\cite{Dziurda_2018}. The threshold values account for the accuracy and precision of the alignment, evaluated using samples of LHCb simulated events generated with Run 2 conditions~\cite{Dujany:2272135}, and compared to the alignment constants variations observed in 2012 and 2015 data for which the numbers are reported in Table \ref{tab:tab1}. 

\begin{table}[ht]
    \centering
    \begin{tabular}{||c c ||} 
        \hline
        d.o.f & Min variation \\ [0.5ex] 
        \hline\hline
         T$_x$, T$_y$ [\mum] & 1.5 \\ 
        \hline
         T$_z$ [\mum] & 5 \\
        \hline
        R$_x$, R$_y$ [\murad] & 4 \\
        \hline
        R$_z$ [\murad] & 30 \\
        \hline
    \end{tabular}
    \caption{Alignment constants thresholds used for the online update~\cite{Dujany:2272135}}
    \label{tab:tab1}
\end{table}

\section{Survey and metrology measurements}
\label{sec:metrology}  

During the construction and assembly of the \velo a series of survey measurements were taken to evaluate the position of the each element of the detector \cite{Huse:1067148,Sutcliffe:1082459}.
This included a Coordinate Measuring Machine (CMM) metrology for each module after its assembly in each VELO half. All these measurements were taken at the ambient temperature, about $25$\degc.
The accuracy of the module position was evaluated to be of about 15\mum, 50\mum, and 200\mum for translations along the $x$, $y$, and $z$ axis, respectively, and of the order of 1~mrad for rotations around the $x$ and $y$ axes and 0.2~mrad for rotations around the $z$ axis.
The positions of the two \velo halves were determined in photogrammetry measurements performed at the final \velo setup in the experimental cavern before insertion of the two detector halves \cite{EDMS-908162}. The accuracy of the single measurement is about 100\mum for the translations and 100\muRad for the rotations around the $x$ and $y$ axes.

These measurements were used as initial position of the detector for the alignment procedure based on the reconstructed tracks. The track-based alignment results have a higher accuracy. The accuracy of the survey measurements was expected to be comparable with the difference between the survey measurements and the alignment results. A variation of the alignment of the detector halves up to 10\mum between fills is expected, due to the precision of the resolvers as part of the motion system \cite{lhcb-exp}. 
This applies to the alignment of the detector halves, not to modules within the halves. Instead the alignment of the two halves showed a significant misalignment along the $x$ axis of about 160\mum for each half, beyond the accuracy of the survey measurements.
This corresponds to a distance of about 320\mum between the 2 halves at the nominal closed position of the \velo. The track efficiency of the detector was not affected by this opening because in its nominal position the acceptance of the two \velo halves overlaps by about 1.7~mm in $x$.

\begin{figure}[H]
\centering
\begin{subfigure}{.5\textwidth}
  \centering
  \includegraphics[width=1.0\linewidth]{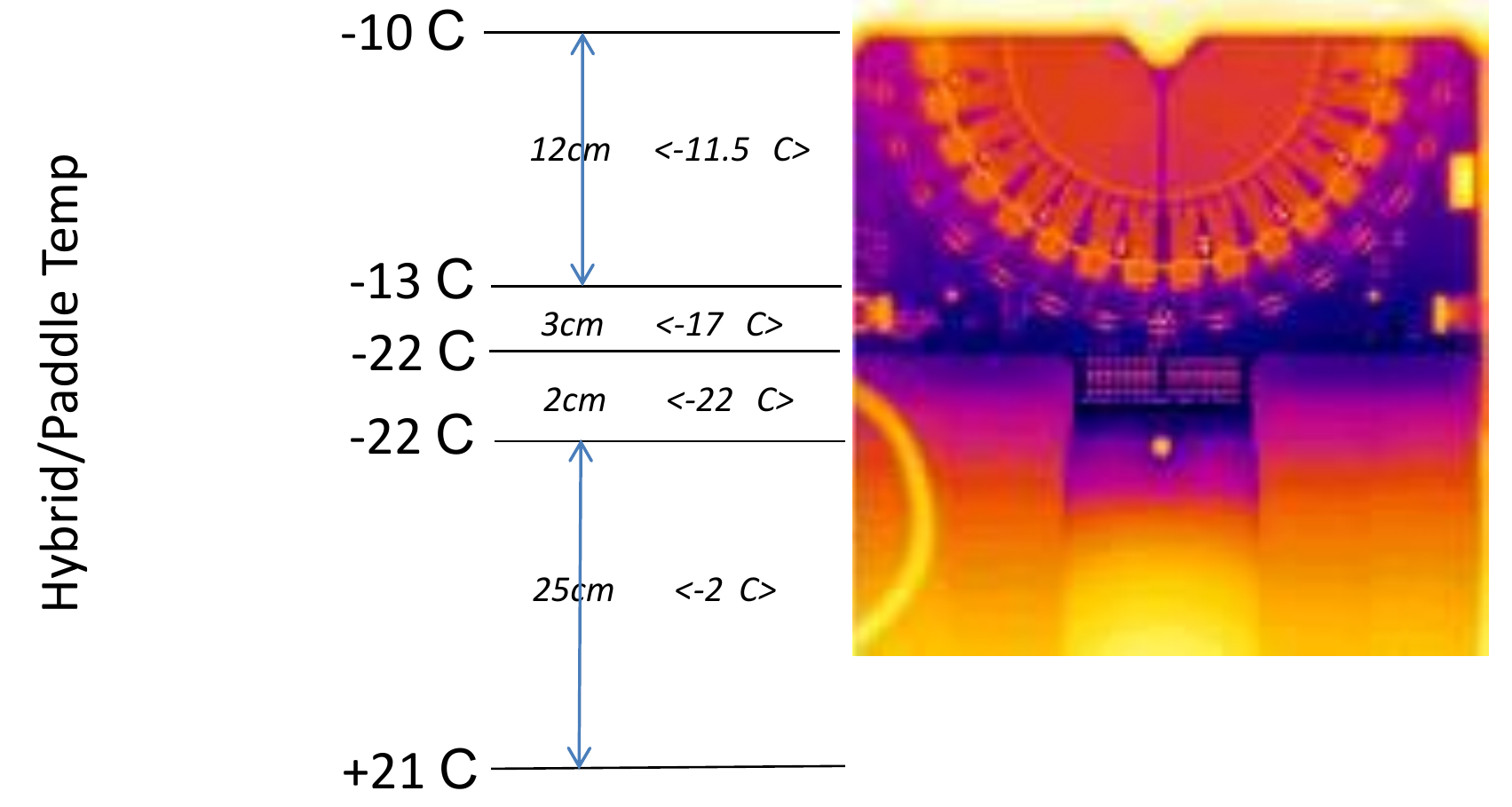}
  \caption{}
  \label{fig:velo-temperatureisothermal}
\end{subfigure}%
\begin{subfigure}{.5\textwidth}
  \centering
  \includegraphics[width=0.8\linewidth]{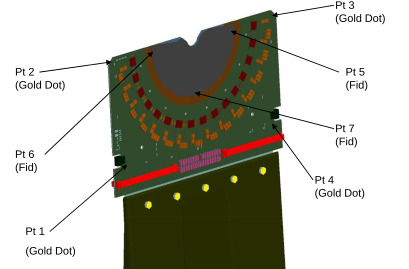} 
  \caption{}
    \label{fig:velo_metrologypoints}
\end{subfigure}
\caption{Temperature gradient along the silicon and the hybrid due to the to isothermal effect, in the case of a cooling temperature at $-25$\degc (a). A drawing of the \velo module illustrating the approximate positions of the survey points (b).}
\end{figure}

The survey campaigns were taken at the ambient temperature, while the operation temperature of the \velo is $-30$\degc. To investigate if the cause of these displacements were due to the temperature difference between the survey campaign and the operation condition, CMM metrology on a single spare module was taken at different temperatures in the laboratory.
It should be considered that there is a temperature gradient along the silicon and the hybrid due to the to isothermal effect, shown in Fig.~\ref{fig:velo-temperatureisothermal} as illustrative example in the case of a cooling temperature at $-25$\degc.
Fig.~\ref{fig:velo_metrologypoints} shows the different points measured in this campaign.
The measurements were taken at the following cooling temperature: $+45$\degc, $+20$\degc, $-15$\degc, $-25$\degc.
The position was studied as a function of the cooling temperature and the shrinkage at $-30$\degc was extrapolated by a linear fit on the measurements.

At the operational temperature of $-30$\degc, the shrinkage of the silicon sensor and the hybrid is -31\mum.
This corresponds to about 1\mum per K temperature variation. The main responsible of the shrinkage is the Cu present in the hybrid. This explains an opening distance between the 2 halves for about 70\mum.

\section{\velo Temperature Scan}

At the end of Run 2, a temperature scan of the \velo was performed to evaluate the temperature effect on the detector position. During one of the last fills of the \lhc Run 2, in October 2018, the operating temperature of the detector was changed from $-30$\degc to $-20$\degc using two intermediate steps with temperatures $-26$\degc and $-24$\degc. Before reaching the stable temperature of $-24$\degc the high voltage was changed from 350~V to 250~V. To avoid effects due to the accuracy of the closing system the data were collected in the same fill. In order to reach thermal stability the data was collected 5 minutes after the temperature change. The data samples were saved at each temperature point, and the samples are used to study the effect on the detector performance. In the next section we present the study of the detector movements as a function of the temperature.

\subsection{Results} 

The alignment procedure was run to evaluate the detector element positions (position of each module and of the halves) at each temperature. The variation of the position is evaluated as the variation of the alignment constants with respect to the constants at the \velo operation temperature ($-30$\degc). These variations are analysed as a function of temperature and a linear dependency is expected. To disentangle any variation due to the alignment procedure, the results obtained with different constraints are compared.
The alignment of the modules is performed together with the 2 half alignment. The modules are aligned for the main d.o.f., T$_x$, T$_y$ and R$_z$, while the halves are aligned for all d.o.f.: the $x$, $y$, $z$ translations and rotations. The alignment uses as constraint the modules 10, 11, 32, 33 (that are not aligned) and the average position of the two \velo halves. Module 10, 32 are mounted on the C-side(right) of the \velo and 11, 33 are at the A-side(left). Different constraints are considered in the systematic check.

\subsection{Halves alignment}

\begin{figure}[!b]
\centering
\begin{subfigure}{.5\textwidth}
  \centering 
  \includegraphics[scale=0.4]{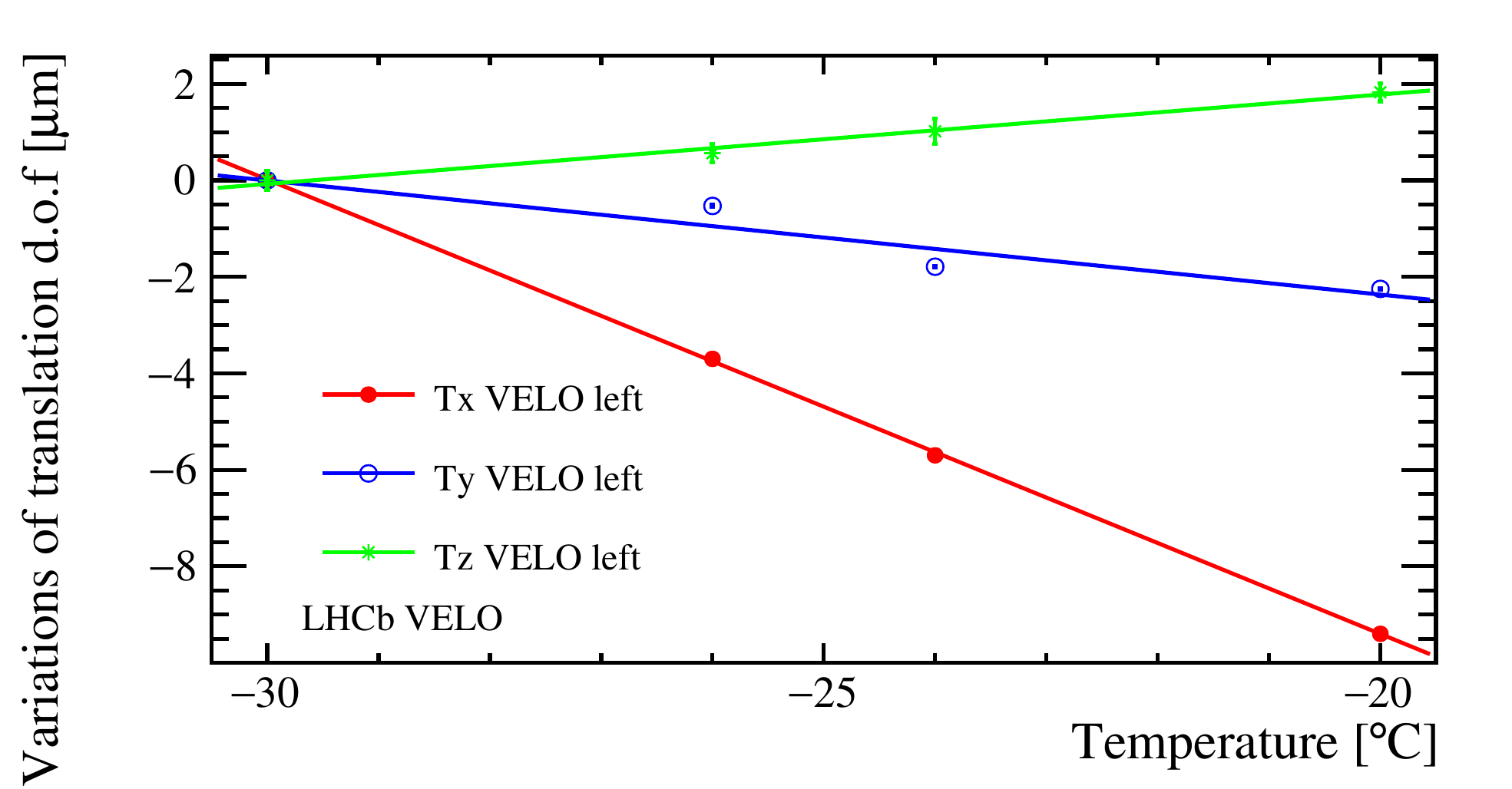}
  \caption{}
\end{subfigure}%
\begin{subfigure}{.5\textwidth}
  \centering
  \includegraphics[scale=0.4]{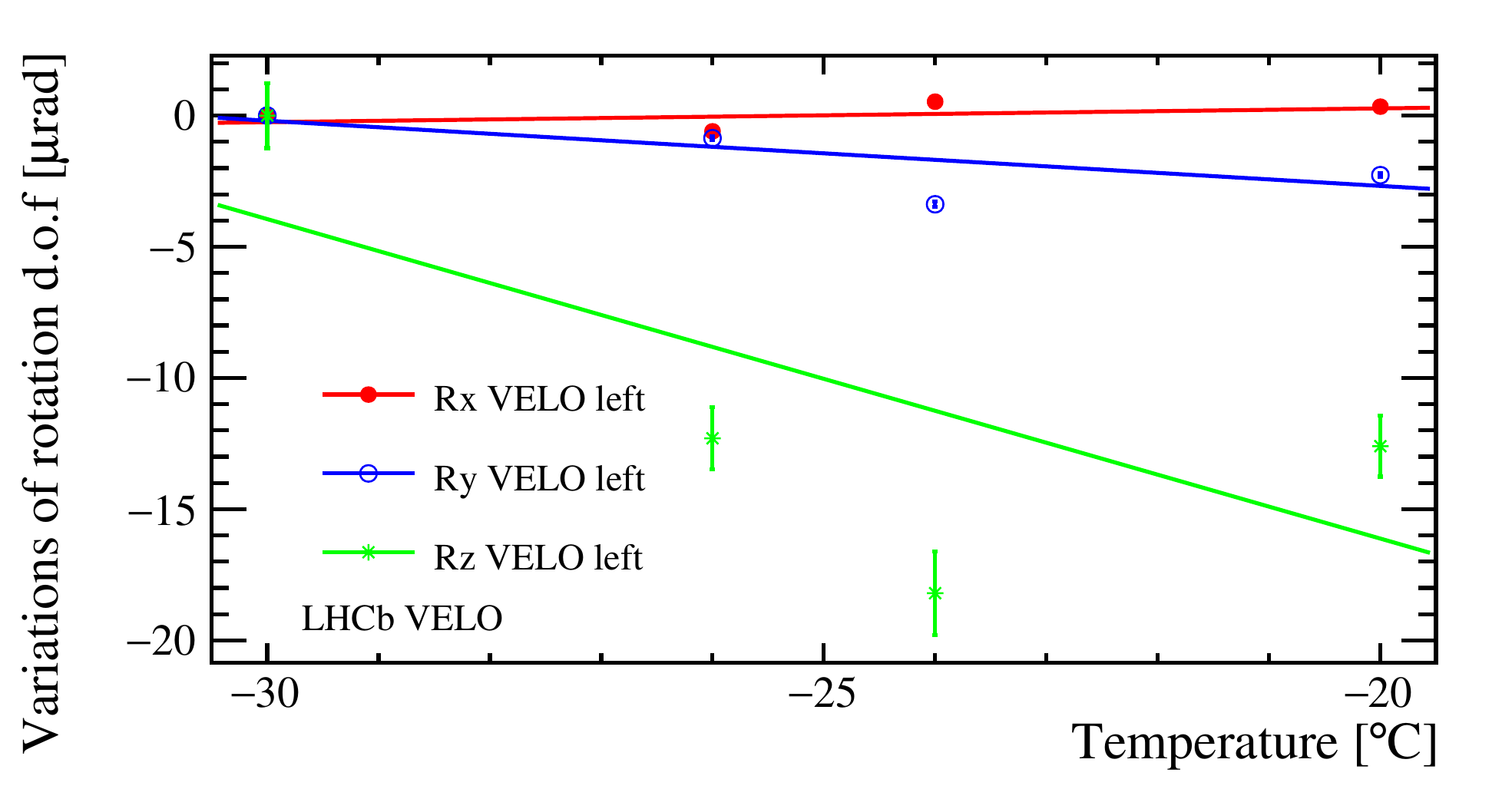} 
  \caption{} 
\end{subfigure}
\caption{Variations of translation (a) and rotation (b) d.o.f of the left \velo, calculated when two halves and modules alignment is performed. The dashed grey line in the left plot is the min variation needed for an update in T$_x$.}
\label{fig:halves-trans-mod}
\end{figure}

On Fig.~\ref{fig:halves-trans-mod}, the variations of the translations in all directions ($x$, $y$ and $z$) for the \velo halves are shown as a function of temperature. The halves are identical per construction and the same movement is expected. The alignment constants for the \velo right half take the same values but opposite sign and the plots that show the temperature dependence are symmetrical.  Therefore, we report the results of only one half (\velo left). A linear trend is visible for all degrees of freedom, indicating higher temperature leads to larger variations. The largest variation is for T$_x$ up to 10\mum and all the rest of the points exhibit a variation larger than accuracy of the alignment procedure \cite{Dujany:2272135}. The variations of T$_y$ and T$_z$ are up to 3\mum and considered almost negligible as they are comparable with the alignment accuracy (Table \ref{tab:tab1}). Table \ref{tab:5} shows the linear fit results for the translation variation along $x$ for the \velo halves as a function of temperature. It results a variation of 0.94\mum per K.
Fig.~\ref{fig:halves-trans-mod} shows the variations of the rotations as a function of temperature. The variations are in the range between -20\muRad and +20\murad for R$_z$ and up to 5\murad for R$_x$ and R$_y$. All the variations are compatible with the alignment accuracy from Table \ref{tab:tab1} and hence they are not significant. 

\begin{table}[t]
    \centering
    \begin{tabular}{||c c c c||} 
        \hline
         \velo Left & \multicolumn{2}{c}{$T_x = p_0 + p_1 T$} & \\ [0.5ex]
        \hline\hline
         Alignment constant & $p_0$ [\mum] &  $p_1$ [\mum/K] & $\chi^2/\rm d.o.f$ \\ 
        \hline
         T$_x$ & \small $\rm -28.24 \pm 0.06$ & \small $\rm -0.942 \pm 0.003$ & 3.9\\
        \hline
    \end{tabular}
    \caption{Linear fit results for variations of T$_x$ alignment constants aligning for halves and modules of the left half of the \velo.}
    \label{tab:5}
\end{table}

\subsection{Modules alignment}

The shrinkage of modules as function of the temperature is expected to be similar for all the modules and the variations are expected to be significantly smaller than the average shrinkage of all the modules included in the the half alignment constants.
The observed variation for the translation along the $x$ direction for the modules has a linear behaviour as a function of the temperature and is a factor of $\sim 3$ smaller than the half T$_x$ variation.
The variations in T$_y$ are smaller than 1\mum, except for the first two modules for which the variation can be as large as 2\mum. Given that the alignment accuracy is 1.5\mum, these variations are considered negligible. The R$_z$ variation of the modules in the left and the right \velo half is up to 60\murad and most of the points are below the accuracy of 30\murad. 
One can evaluate the module position in the global reference system adding the evaluated movement of the half that is obtained by the 2 half alignment. These variations are analysed as function of the $z$ position of the modules. Within this we can study any eventual dependency of the shrinkage on the $z$-position of the module. The 2 half alignment represents the average movement of the modules, while the module alignment is an overall correction on the single module behaviour.  
\begin{figure}[H]
\begin{center}
    \includegraphics[scale=0.5]{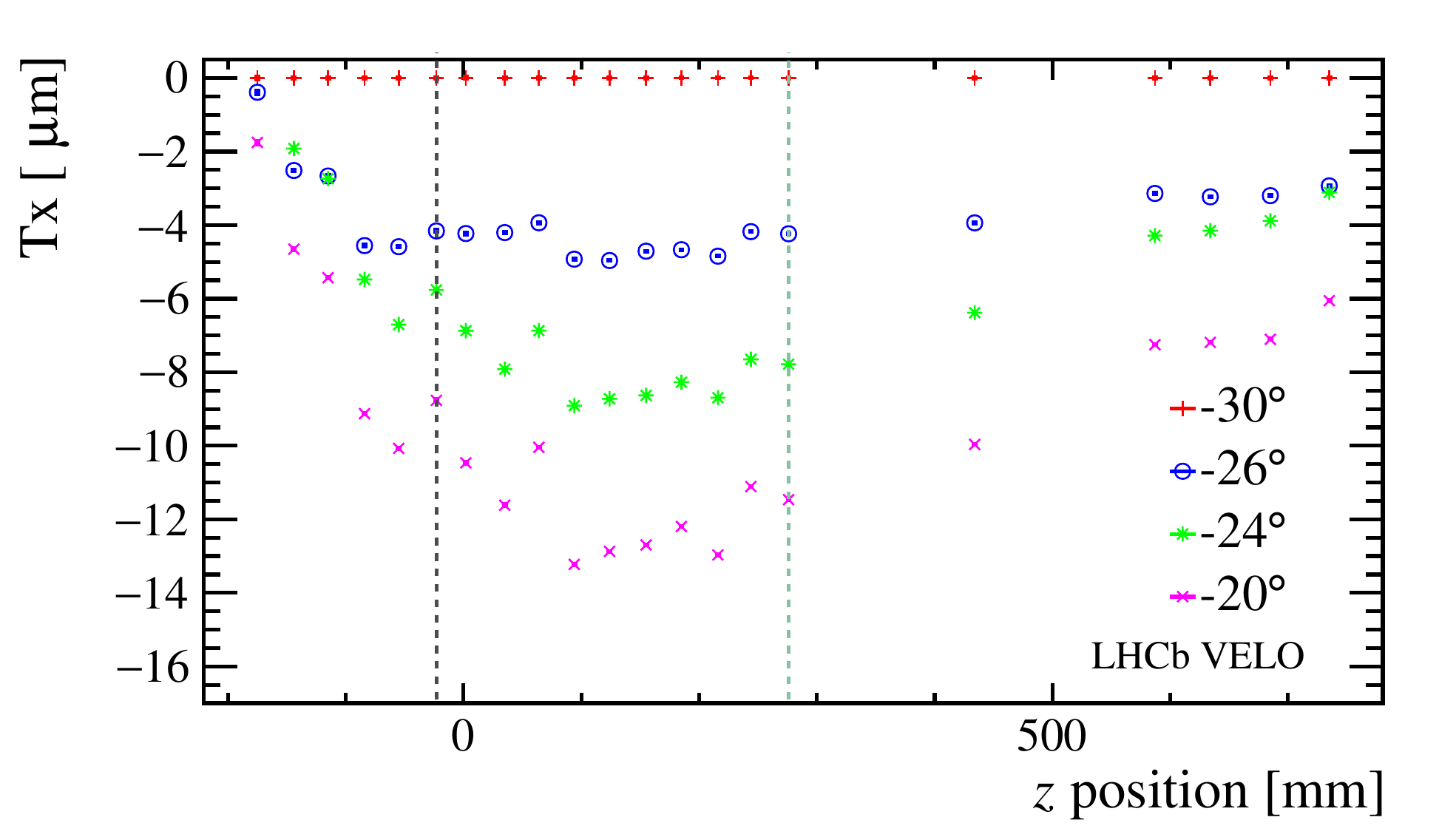}
    \caption{Variations of T$_x$ d.o.f. of the \velo modules of the left \velo half, in the global reference frame as a function of $z$ position of modules. The grey and green vertical lines correspond to the positions of modules 11 and 33, respectively.}
  \label{fig:mod-tx-zpos-left-global}
\end{center}
\end{figure}

Fig.~\ref{fig:mod-tx-zpos-left-global} shows the variations for the modules of the left \velo half as a function of their position in the global frame. One can see a main movement due to the temperature change, up to 12\mum for a 10\degc temperature variation (Fig.~\ref{fig:mod-tx-zpos-left-global}). The other \velo half has a similar behaviour. The variations among modules have a negligible dependence as a function of their $z$ position, except in the backward region, where the variation is up to 6\mum. This variation could be due to the fact that this region is less constrained in the alignment procedure. Due to the fact that halves and modules were aligned simultaneously, with the survey constraints fixing the relative alignment, the presented numbers show a few micron variation depending on how the global weak modes were fixed.

\section{Comparison with previous temperature studies}

\begin{figure}[H]
\begin{center}
    \includegraphics[width=0.7\linewidth]{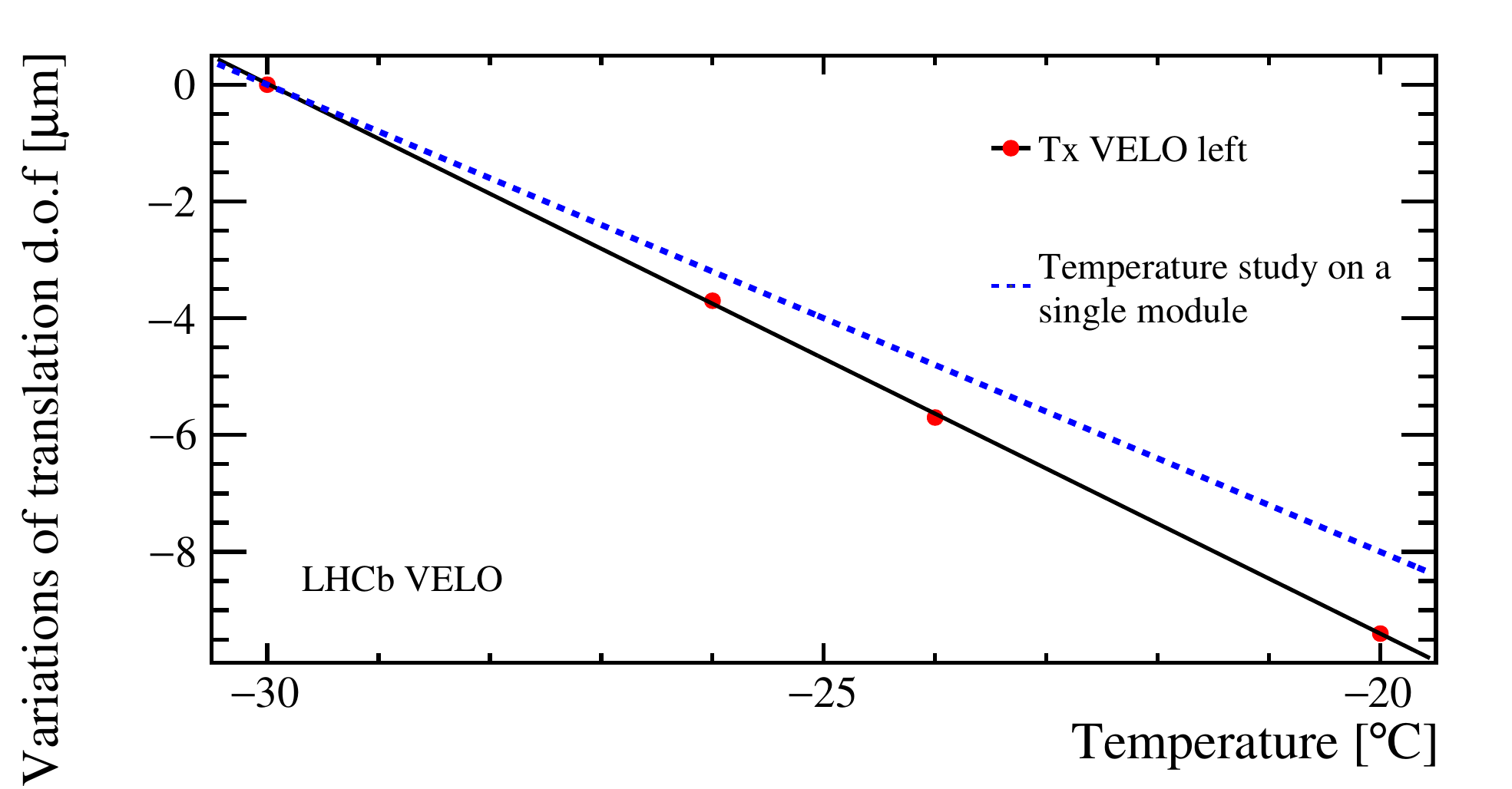}
     \caption{Variations of $\rm T_x$ as a function of temperature (alignment using halves and modules) compared with the study at the single module in laboratory (blue dashed line on the plot).}
     \label{fig:halves-mod-trans-liv}
\end{center}
\end{figure}

The laboratory measurements on a single module can be compared to the half $x$ position variation evaluated by the alignment procedure at each temperature (Fig.~\ref{fig:halves-mod-trans-liv}).

Fig. \ref{fig:halves-mod-trans-liv} shows the modules variations of T$_x$ as function of temperature for the alignment performed for both the halves and the modules. The measurements on a single module are shown in blue. The two results are in good agreement. This confirms the thermal expansion of the \velo modules is as expected from laboratory measurements. By a linear extrapolation, this results a shrinkage of about 30\mum for each half  due to the difference of the temperature during the metrology campaign and the operation campaign. This explains about one third of the observed 320\mum distance between the two halves when the \velo is in the nominal closing position. The other part can be explained by a $0.6\%$ calibration of the motion system along the $x$ direction \cite{LHCb-DP-2014-001}. This was evaluated thanks to the tracking alignment evaluated at different opening position of the \velo halves with the first collision data.

\section{Conclusion}
The thermal properties of the \velo modules are studied using the alignment. Data samples  were collected at four operating temperatures between $-30$\degc and $-20$\degc and the alignment procedure was used to evaluate the \velo alignment constants variations as a function of the temperature. 
Both \velo halves and modules have been aligned for translations and rotations at each temperature point. 
The alignment constants are found to vary linearly as function of temperature. The only significant variation is for the translations along the $x$ axis. 
The modules variations have a nonlinear behaviour with small variations within 2\mum at the same temperature. The variations in T$_y$ and R$_z$ are not significant as they are within the alignment accuracy.
These results are compared with the measurements performed on a single module in laboratory conditions. The variation observed for T$_x$ is compatible with those measurements. 
The two results are in good agreement and confirm a shrinkage of 1\mum per K. The difference between the temperature at which the survey measurements were performed ($+25$\degc) and the operation temperature of the detector ($-30$\degc) leads to a thermal shrinkage of about 30\mum for each half. 
\newpage 


\addcontentsline{toc}{section}{References}
\bibliographystyle{LHCb}
\bibliography{main,standard,LHCb-PAPER,LHCb-CONF,LHCb-DP,LHCb-TDR}
 
\end{document}